\documentclass{article}
\usepackage{iclr2027_conference,times}

\usepackage{amsmath,amsfonts,bm}

\def\eqref#1{equation~\ref{#1}}

\def\1{\bm{1}}

\DeclareMathAlphabet{\mathsfit}{\encodingdefault}{\sfdefault}{m}{sl}
\SetMathAlphabet{\mathsfit}{bold}{\encodingdefault}{\sfdefault}{bx}{n}

\usepackage{hyperref}
\usepackage{url}

\usepackage{graphicx}
\usepackage{listings}
\usepackage{xcolor}
\usepackage{booktabs}
\usepackage{enumitem}
\usepackage{array}
\usepackage{algorithm}
\usepackage{algpseudocode}
\usepackage{enumitem}
\usepackage{subcaption}
\usepackage{tabularx}
\usepackage{array}
\newcolumntype{L}[1]{>{\raggedright\arraybackslash}p{#1}}

\graphicspath{{figures/}}

\definecolor{addition}{rgb}{0, 0.6, 0}   
\definecolor{deletion}{rgb}{0.8, 0, 0}   
\definecolor{unchanged}{rgb}{0, 0, 0}    
\definecolor{skillbg}{RGB}{250,249,244}  

\lstdefinelanguage{Jac}{
  morekeywords={walker,node,edge,obj,enum,has,can,with,entry,exit,visit,disengage,report,spawn,def,sem,glob,import,from,by,llm,impl,if,else,elif,for,while,return,try,except,and,or,not,in,True,False,None,str,int,bool,list,dict,float,self},
  sensitive=true,
  morecomment=[l]{\#},
  morecomment=[l]{//},
  morestring=[b]",
  morestring=[b]'
}

\lstdefinelanguage{SkillMD}{
  morekeywords={must,never,always,before,identify,run,read,verify,no,without,require,check,block,escalate},
  sensitive=false,
  morecomment=[l]{\#\#}
}

\lstdefinestyle{paperlst}{
  columns=fullflexible,
  breaklines=true,
  captionpos=t,
  basicstyle=\fontsize{8}{9}\selectfont\upshape\ttfamily,
  numbers=left,
  numbersep=8pt,
  numberstyle=\tiny\color{gray},
  keywordstyle=\bfseries\color{blue!55!black},
  stringstyle=\color{orange},
  commentstyle=\color{gray}\ttfamily,
  xleftmargin=5pt,
  xrightmargin=10pt,
  backgroundcolor=\color{white},
  morekeywords={self},
  emph={return,by,py,jac,entry,type,str,int,float,list,dict,bool},
  emphstyle=\bfseries\color{green!50!black},
  frame=tb,
  showstringspaces=false,
  keepspaces=true
}

\lstdefinestyle{skillfig}{
  style=paperlst,
  language=SkillMD,
  numbers=none,
  basicstyle=\fontsize{7}{8}\selectfont\upshape\ttfamily,
  keywordstyle=\bfseries\color{blue!55!black},
  backgroundcolor=\color{skillbg},
  xleftmargin=3pt,
  xrightmargin=3pt,
  aboveskip=3pt,
  belowskip=3pt,
  frame=single,
  rulecolor=\color{black!20},
  framesep=5pt,
  showstringspaces=false,
  keepspaces=true
}

\newcommand\jacinline[1]{{\lstset{language=Jac}\lstinline!#1!}}

\providecommand{\Description}[1]{}

\title{SIGIL: Skill Compilation for Reliable and \\Efficient Agent Execution}

\author{Jayanaka Dantanarayana, Savini Kashmira, Lingjia Tang \& Jason Mars \\
University of Michigan, Ann Arbor, USA \\
\texttt{\{jayanaka,savinik,lingjia,profmars\}@umich.edu}}

\iclrfinalcopy 

\begin{document}

\maketitle
\lhead{} 

\begin{abstract}
Agent skills describe reusable procedures, but runtime models
must still interpret their instructions and coordinate execution.
We introduce \textbf{Skill Compilation}, which translates the
procedure prescribed by an authored skill into an executable
harness while preserving decisions left to the model.
Our compiler, \textbf{SIGIL}, translates skills and their resources
into a typed intermediate representation, validates it, and
deterministically generates the harness.
We evaluate SIGIL on 11 compliance-critical skills, where following
the prescribed procedure is part of correctness.
SIGIL improves adherence across all four runtime models, achieving
up to 100\% measured mean Skill Adherence compared with
26.2--54.0\% for direct \texttt{SKILL.md} execution.
On this suite, SIGIL reduces total runtime tokens by 21--45\%
for three of the four models.
On SkillsBench, SIGIL also improves task performance across all
four models, with absolute gains of 3.5–35.7 percentage points.
These findings suggest that compiling reusable skill procedures
into executable harnesses can improve adherence, task completion,
and runtime efficiency.
\end{abstract}

\section{Introduction}
\label{sec:intro}

Large Language Model (LLM)-driven agents are increasingly used for tasks
that require domain-specific tools, procedures, and knowledge. Many of
these procedures recur across queries and applications. Agent skills
package procedural instructions, scripts, and supporting resources into
reusable modules. Formats such as \texttt{SKILL.md} can specify which
tools to use, which steps to follow, how actions should be ordered, how
exceptions should be handled, and what actions must not
occur~\citep{anthropic2025skills,agentskills2025spec}. Skills can be
shared and deployed across agents, contributing to a rapidly growing
public ecosystem. Recent analysis of more than 100,000 public skills
finds that most provide tool-use or procedural guidance, often with
ordered steps, branches, and data dependencies~\citep{skillrt2026}.
Skills are therefore emerging as a natural-language interface for
specifying reusable agent procedures.

Providing a skill to an agent does not ensure that the agent will follow
its instructions exactly as specified. The runtime model must interpret
the prescribed steps, their ordering, and stopping conditions while
coordinating execution and making the semantic decisions required by
the task. Execution can therefore vary with instruction wording, model
choice, and across several runs~\citep{lu2022fantastically}. The model
may skip required steps, perform them in the wrong order, or continue
past a specified stopping point, failing to adhere to the skill's
procedure. These errors can prevent the agent from completing the task
successfully. Even when the agent follows the procedure correctly,
repeatedly interpreting the skill and deciding which step to take next
can increase token use and execution cost, although the skill already
specifies the required steps and their order. This motivates an approach
that improves skill adherence and task performance while reducing the
cost of procedural coordination.

\begin{figure*}[t]
    \centering

    \begin{subfigure}[c]{0.60\textwidth}
        \centering
        \includegraphics[width=\linewidth]{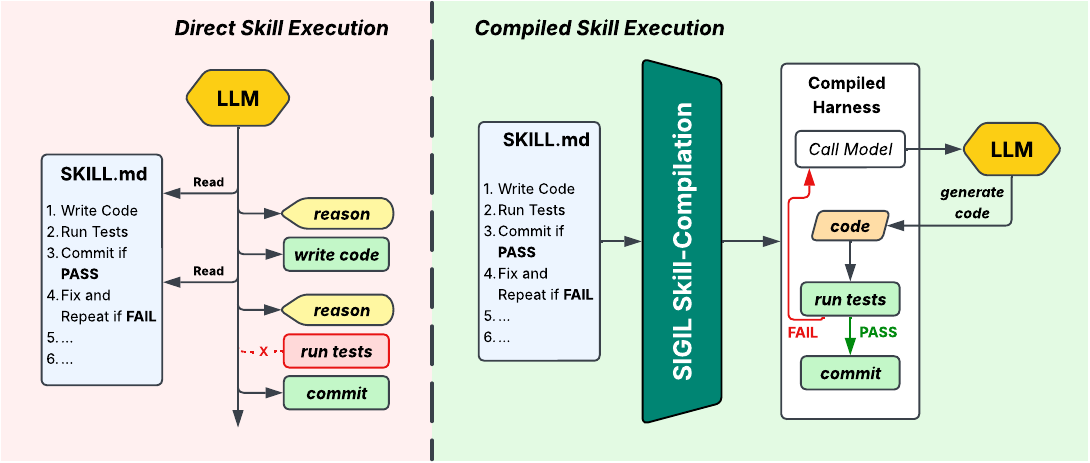}
        \caption{}
        \label{fig:sigil-overview}
    \end{subfigure}
    \hfill
    \begin{subfigure}[c]{0.39\textwidth}
        \centering
        \includegraphics[width=\linewidth]{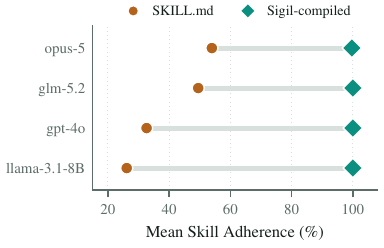}
        \caption{}
        \label{fig:sigil-highlight}
    \end{subfigure}

    \caption{
    (a) SIGIL compiles natural-language skills into executable agent
    programs, encoding prescribed procedure in program structure while
    retaining semantic model operations.
    (b) Harnesses compiled using Claude Opus 5 achieve 100\% measured
adherence across all four runtime models, versus 26.2--54.0\%
for \texttt{SKILL.md}.
    }
    \label{fig:sigil-teaser}
    \vspace{-0.5cm}
\end{figure*}

Existing approaches improve skill execution through instruction
optimization and adaptation to the execution environment. GEPA revises
instructions using execution feedback~\citep{agrawal2025gepa}, while
SkVM adapts skills to the capabilities of the target model and
harness~\citep{skillrt2026}. However, better instructions alone do not
ensure that the model follows the prescribed procedure. To enforce requirements during execution, runtime guards check proposed actions against rules and block violations~\citep{wang2026agentspec}. These checks can prevent prohibited actions, but checking proposed actions alone does not ensure that every required step is performed. To control which steps run and in what order, developers can encode the procedure in a hand-written harness, retaining model judgment where needed. Doing this manually requires developers to identify required steps, dependencies, and exceptions, determine which parts need model judgment, and encode them without omitting requirements or changing their meaning. They must also keep the harness consistent with later skill revisions~\citep{balis2026workflow}. This motivates automatically deriving a harness from an authored skill while preserving its required operations, ordering, conditions, and prohibitions, and retaining the model judgment and delegated execution the skill leaves open. 


\textbf{Our key insight is that an agent harness can enforce the skill's
required steps and ordering, while the model handles the semantic
decisions required by the task.} Recovering the required steps and
dependencies makes the procedure explicit, but following that procedure
still requires coordinating its execution. A reusable harness can take
on this coordination, leaving semantic decisions and open subprocedures
to the runtime model. This separation can improve skill adherence and
task performance by reducing procedural errors, while lowering the cost
of repeated interpretation and coordination.

Based on this insight, we introduce \textbf{Skill Compilation} and realize
it in \textbf{SIGIL}, a compiler that translates natural-language skills
into executable agent harnesses. A compile-time agent explicitly extracts
source-grounded requirements from the skill, checks for missing
requirements, and revises the extracted specification. It also identifies
the tool bindings and knowledge needed for execution. We define the
\textbf{Agent Instruction Set (AIS)}, a set of basic operations for model
decisions, tool use, and control flow. The compile-time agent represents
the extracted requirements using these operations and organizes them into
an \textbf{Agent Intermediate Representation (AG-IR)}. AG-IR makes
execution modes, typed data dependencies, and control flow explicit,
and connects operations to the tools and knowledge they need. SIGIL
checks AG-IR against the extracted requirements and validates its
structural consistency. It then deterministically generates an executable
harness without reinterpreting the skill. A capable model can compile
the skill once, and the resulting harness can be reused across runtime
models. SIGIL generates harnesses in Jac~\citep{jaclang}, a programming
language with support for both procedural control flow and typed model
calls. These features let SIGIL easily translate AG-IR's control flow
and model operations into executable programs.

We evaluate SIGIL on 11 public compliance-critical skills~\citep{riskresponse_grc_skills,naik2026grcskills,kassis2026scientificagentskills} that provide sufficient guidance for end-to-end execution, where following the prescribed procedure is part of correctness. Across four runtime models, SIGIL harnesses compiled with Claude Opus 5 achieve 100\% measured mean Skill Adherence, while those compiled with GLM-5.3 and GPT-5.6-Sol achieve 96.4--100\%. These results compare with 26.2--54.0\% when agents execute the original \texttt{SKILL.md} directly. On the same suite, Opus-compiled harnesses reduce average tokens per LLM call across all four runtime models. Total runtime tokens decrease by 21--45\% for three models, excluding compilation cost; Llama-3.1-8B uses 2.13 times as many total tokens while completing previously omitted steps. On SkillsBench~\citep{skillsbench}, SIGIL improves task performance, with Task Macro Pass Rate increasing by 20.8 percentage points for Claude Opus 5, 35.7 for GPT-4o, 34.1 for Llama-3.1-8B, and 3.5 for GLM-5.2.

The key contributions of this paper are:
\begin{enumerate}[nosep]
    \item \textbf{Skill Compilation} and \textbf{SIGIL}, an approach
    and compiler that turn natural-language skills into executable
    harnesses that enforce prescribed procedures while retaining
    runtime model judgment;

    \item \textbf{AIS} and \textbf{AG-IR}, an instruction set and
    intermediate representation that specify operations, execution
    modes, control flow, data dependencies, and links to source
    requirements, tools, and knowledge; and

\item An \textbf{evaluation} showing that SIGIL improves skill
    adherence, reduces total runtime token use, and improves task performance
    over agents given the original \texttt{SKILL.md} directly.
\end{enumerate}

\section{Problem Formulation and Skill-Compilation Requirements}
\label{sec:motivation}

\begin{figure}[t]
    \centering
    \begin{minipage}[t]{0.74\columnwidth}
\begin{lstlisting}[
    style=skillfig,
    aboveskip=0pt,
    belowskip=0pt
]
24  Is this data PHI?
...
26  Does a vendor or model provider require a BAA
...
28  [check that read/write/export events are auditable]

31  Never place PHI in logs
...
33  [require authenticated, scoped, auditable PHI access]
34  [block third-party services until BAA status and
     data boundaries are established]
...
43  [request: add AI-generated visit summaries]
...
49  ... before any PHI is sent
50  [escalate if summaries affect clinical decisions]
\end{lstlisting}
    \end{minipage}
    \hfill
    \begin{minipage}[t]{0.205\columnwidth}
        \vspace{0pt}
        \centering
        \includegraphics[width=\linewidth]{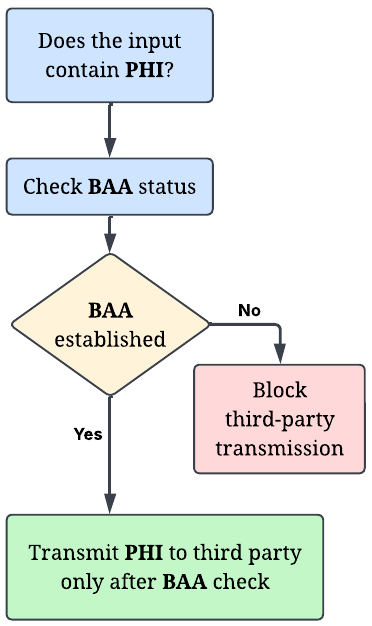}
    \end{minipage}

    \caption{
        Excerpt from a public \texttt{hipaa-compliance}
        skill~\citep{naik2026grcskills} and a simplified PHI transmission
        path where the skill requires a BAA check.
    }
    \label{fig:hipaa-skill}
\end{figure}

A natural-language skill specifies operations and relationships among
them. We use a public \texttt{hipaa-compliance}
skill~\citep{ecc2026hipaa} for healthcare workflows to distinguish
prescribed procedural structure from runtime model judgment and define
what Skill Compilation must preserve.

\subsection{From Skill Text to an Agent Procedure}
\label{sec:motivation-gap}

Figure~\ref{fig:hipaa-skill} illustrates how a skill combines semantic
decisions with procedural requirements. The agent must identify whether
the input contains protected health information (PHI; line 24). It must
also establish the applicable Business Associate Agreement (BAA) status
before transmitting PHI to a third party (lines 26, 34, and 49).
Identifying PHI may require model judgment, while the required BAA check
and its ordering before transmission can be enforced by a harness.

When the skill is provided as text, the model must both identify PHI
and follow the prescribed ordering. Recognizing that the BAA check
must precede transmission does not ensure that the model performs
that check before transmitting. A harness can make the check part
of the execution path and block transmission unless the applicable
condition is satisfied, while leaving PHI identification to the model.
This enforces the required check and ordering, but does not guarantee
correct PHI identification.

\subsection{Procedural Structure and Operations}
\label{sec:motivation-separation}

The example above distinguishes operations from the rules specifying
their order and the conditions under which they must execute.
We call these rules \emph{procedural structure}: the ordering,
dependencies, branches, loops, parallel execution, and prohibited paths
connecting operations and subprocedures. The harness must enforce this
structure while preserving the skill's instructions for performing each
operation or subprocedure. These instructions, together with the
available resources, determine which of the following three execution
types applies.

\begin{enumerate}[nosep]
    \item \textbf{Executable operations.}
    When the skill or its resources provide a function, script, database
    query, or tool for an operation, the harness calls that implementation.
    For example, if a tool for checking BAA status is provided, the harness
    uses it to perform the required check.

    \item \textbf{Semantic computations.}
    When an operation requires interpretation and no task-specific
    implementation is provided, the harness asks the model to produce its
    result, such as identifying whether an input contains PHI. The harness
    determines the operation's inputs and its place in the procedure, while
    the result depends on model judgment.

    \item \textbf{Delegated execution.}
    The harness assigns a subprocedure to a runtime agent when its internal
    steps are left open. For example, the HIPAA skill requires escalation
    to \texttt{healthcare-reviewer} when visit summaries influence clinical
    decisions (line 50). The harness can enforce when this review is
    requested, while the reviewer uses model judgment and available tools
    to choose and carry out any unspecified review steps within the
    skill's constraints.
\end{enumerate}

\subsection{Skill Compilation and Its Requirements}
\label{sec:motivation-requirements}

\emph{Skill Compilation} recovers procedural structure and execution
types from the skill and its resources, then represents them in an
executable agent harness. The challenge is to avoid omitting required
steps, misinterpreting exceptions, or changing how operations are
performed. Faithful compilation must therefore satisfy the following
requirements:

\begin{enumerate}[nosep]
    \item \textbf{Source fidelity.}
    Preserve required, discretionary, and prohibited behavior,
    including applicable conditions and exceptions, without
    introducing unsupported requirements. Retain links from
    represented requirements to their source text.

    \item \textbf{Structural realization.}
    Make specified ordering, dependencies, and conditions explicit
    in program structure. Applicable required operations must
    execute on the relevant paths, while prohibited behavior must
    be excluded. For example, the transmission path in
    Figure~\ref{fig:hipaa-skill} must enforce the applicable BAA
    condition.

    \item \textbf{Execution preservation.}
    Preserve each operation's execution type, tools, and supporting
    knowledge. Use the functions, scripts, or tools specified for
    executable operations and retain model judgment for semantic
    computations. Keep open subprocedures delegated without
    decomposing them during compilation.
\end{enumerate}

Together, these requirements define faithful compilation: preserving
the skill's prescribed behavior while retaining the runtime decisions
it leaves open. This guides the design of SIGIL.

\section{SIGIL Skill Compilation}
\label{sec:method}

\begin{figure}[t]
    \centering
    \includegraphics[width=\linewidth]{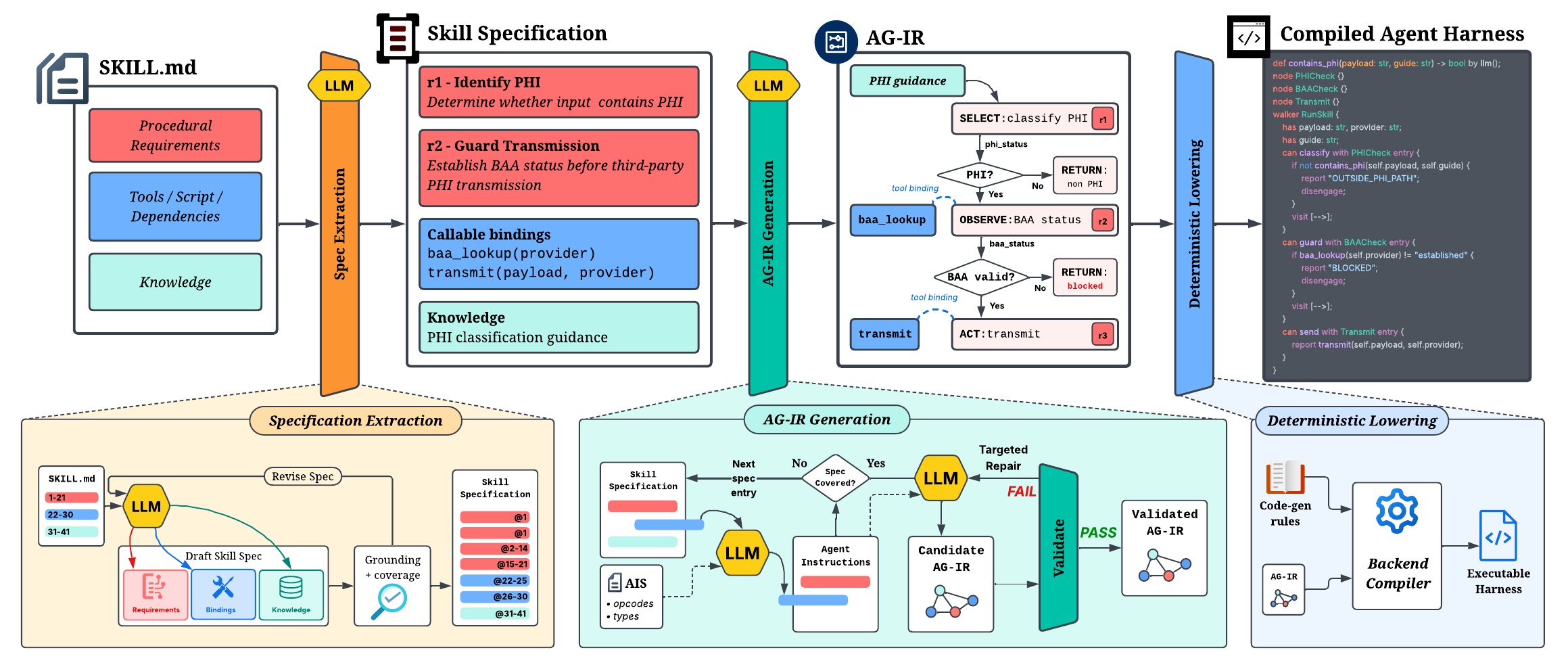}
    \caption{
    SIGIL compilation pipeline: source interpretation and resource preparation,
    AG-IR construction, validation and repair, and deterministic lowering
    into an executable agent harness.
    }
    \label{fig:sigil-pipeline}
\end{figure}

SIGIL compiles a natural-language skill and its resources into an executable
agent harness through the pipeline in Figure~\ref{fig:sigil-pipeline}.
A \emph{compile-time model} recovers the skill's requirements and constructs
AG-IR. SIGIL validates this representation and lowers it deterministically
into a harness.

\subsection{Agent Instruction Set}
\label{sec:ais}

The \textbf{Agent Instruction Set (AIS)} represents an authored skill's
procedural structure, executable operations, semantic computations, and
delegated execution, as defined in Section~\ref{sec:motivation-separation}.

An AIS instruction is $i = \langle o, R, W, \theta \rangle$, where $o$ is
the opcode, $R$ and $W$ reference typed values read and written by the
instruction, and $\theta$ contains parameters such as alternatives, schemas,
bindings, resources, or control-flow cases. The opcode determines execution
semantics. Table~\ref{tab:ais} summarizes the instruction roles. Semantic
instructions leave a result to the model; \texttt{DELEGATE} leaves a bounded
subprocedure's internal orchestration open.

In Figure~\ref{fig:hipaa-skill}, line~24 maps to a semantic selection:

{
\setlength{\abovedisplayskip}{2pt}
\setlength{\belowdisplayskip}{2pt}
\setlength{\abovedisplayshortskip}{0pt}
\setlength{\belowdisplayshortskip}{2pt}
\renewcommand{\arraystretch}{0.85}
\[
\texttt{SELECT(input; \{PHI, NOT\_PHI\})}
\rightarrow
\texttt{phi\_status}.
\]
}

\begin{table}[t]
\caption{The Agent Instruction Set. Instructions are grouped by their role
in the agent procedure.}
\label{tab:ais}
\centering
\small
\resizebox{0.85\linewidth}{!}{%
\begin{tabular}{p{0.10\linewidth} p{0.10\linewidth} p{0.695\linewidth}}
\toprule
\textbf{Class} & \textbf{Instruction} & \textbf{Meaning} \\
\midrule

Semantic
& \texttt{SELECT}
& Choose among a specified set of semantic alternatives. \\

& \texttt{EXTRACT}
& Derive typed or structured information from semantic input. \\

& \texttt{MUTATE}
& Modify an existing value according to semantic instructions. \\

& \texttt{GENERATE}
& Produce new open-ended semantic content. \\

\midrule

Delegated
& \texttt{DELEGATE}
& Execute a bounded subtask whose internal orchestration is deliberately left
to the runtime agent. \\

\midrule

Executable
& \texttt{COMPUTE}
& Invoke an available local computation or implementation. \\

& \texttt{OBSERVE}
& Obtain a value from an external source. \\

& \texttt{ACT}
& Cause an externally visible effect. \\

\midrule

Control
& \texttt{BRANCH}
& Select an explicit successor using an available value or predicate. \\

& \texttt{LOOP}
& Repeat an explicitly represented procedure region. \\

& \texttt{PARALLEL}
& Execute explicitly represented regions independently or concurrently. \\

& \texttt{CALL}
& Invoke another named procedure. \\

\midrule

Termination
& \texttt{RETURN}
& Complete the current procedure and return its result. \\

\bottomrule
\end{tabular}%
}
\vspace{-0.5cm}
\end{table}

If BAA status is available through a registry, obtaining it uses
\texttt{OBSERVE}. AG-IR also represents constraints on relationships
among instructions.

\subsection{Agent Intermediate Representation}
\label{sec:agir}

\textbf{Agent Intermediate Representation (AG-IR)} composes AIS instructions
into a typed control- and data-flow graph that can be validated independently
of a target runtime. AG-IR can be written by a developer or model;
SIGIL generates it from the skill's recovered requirements.
An AG-IR program is $A = \langle I, E, C, K, B, P \rangle$, where $I$
contains AIS instructions, $E$ defines control flow, and $C$ contains named
typed values, called \emph{carries}, connecting instruction outputs to inputs.
$K$ holds instruction-scoped knowledge, $B$ holds executable bindings,
and $P$ links source requirements to their realizations and source spans.

For example, in Figure~\ref{fig:sigil-pipeline}, the PHI \texttt{SELECT}
receives the input and any classification guidance supplied by the skill
through $K$. On the PHI branch, a supplied registry binding in $B$ supports
the BAA \texttt{OBSERVE}, whose result, \texttt{baa\_status}, feeds a
\texttt{BRANCH}:

{
\setlength{\abovedisplayskip}{2pt}
\setlength{\belowdisplayskip}{2pt}
\setlength{\abovedisplayshortskip}{0pt}
\setlength{\belowdisplayshortskip}{2pt}
\renewcommand{\arraystretch}{0.85}

\[
\texttt{BRANCH(baa\_status)}
\rightarrow
\begin{cases}
    \texttt{ACT(payload; ThirdPartyService)}, & \text{valid},\\
    \texttt{RETURN(blocked)},                 & \text{otherwise}.
\end{cases}
\]
}

No executable path reaches this transmission \texttt{ACT} unless
\texttt{baa\_status} is valid. Other requirements constrain data flow:
``never place PHI in logs'' can be represented by excluding represented
paths that supply identified PHI-carrying values to a logging \texttt{ACT}.

\subsection{From Skills to AG-IR}
\label{sec:front-end}

\paragraph{Source interpretation and resource preparation.}
The compile-time model extracts procedural requirements from the skill's
\texttt{SKILL.md} and supporting resources, as shown at the lower left of
Figure~\ref{fig:sigil-pipeline}. For each requirement, it records the source
span, whether it is required, optional, or prohibited, and its applicable
conditions. Tool descriptions are matched to available implementations to
form callable bindings; domain guidance and reference material become
knowledge blocks. Source passages may serve multiple roles without every
line becoming a requirement. A model critic compares the extracted
specification with the source, checking for omitted requirements,
conditions, exceptions, ordering constraints, and prohibitions.
Identified omissions trigger targeted re-extraction.

\paragraph{Requirement-to-AG-IR construction.}
Given these outputs and the AIS specification, the model constructs
instructions, carries, control flow, and constraints that realize the
requirements, as shown at the lower center of
Figure~\ref{fig:sigil-pipeline}. It connects operations to
callable bindings in $B$, attaches knowledge in $K$ where needed, and
retains provenance links. A requirement may map to several instructions
or a constraint on their control or data flow; open subprocedures
use \texttt{DELEGATE}.

\paragraph{Validation and repair.}
In the validation-and-repair loop shown in the same panel, SIGIL checks
requirement realization, type and control-flow consistency, binding and
knowledge references, provenance, and represented ordering and path
constraints. Required executable bindings must resolve before lowering.
Mechanical defects are repaired directly when possible;
interpretation-dependent failures return to the model with relevant source
and AG-IR context. Compilation fails if errors remain after the repair budget
is exhausted. Structural validation establishes internal consistency;
correct source interpretation and requirement realization still depend on
semantic judgment. Appendix~\ref{app:method} provides the compilation
algorithm and validation checks.

\subsection{Deterministic Lowering and Runtime Execution}
\label{sec:lowering}

Validated AG-IR is lowered to Jac using fixed rules, without model calls or
source reinterpretation (Figure~\ref{fig:sigil-pipeline}). Semantic
instructions lower to typed \texttt{byLLM}
operations~\citep{dantanarayana2025mtp}. \texttt{DELEGATE} lowers to a bounded
agentic subprocedure with its declared context, tools, and output contract.
\texttt{COMPUTE} lowers to local program logic, while \texttt{OBSERVE} and
\texttt{ACT} lower to their bound external operations. Control instructions
lower to Jac control flow, and carries become typed runtime state.

At runtime, the harness passes state between operations, invokes bound
implementations, and supplies model calls with their declared inputs and
scoped knowledge. Returned values determine subsequent execution. Skills
aggregated before compilation can produce one artifact; separately compiled
artifacts are selected and coordinated by an outer runtime agent.

\section{Evaluation}
\label{sec:eval}

We evaluate Skill Compilation through three research questions:

\begin{enumerate}[label=\textbf{RQ\arabic*.}, leftmargin=*, nosep]

\item \label{rq:compliance}
\textbf{Does Skill Compilation improve adherence to skill requirements?}

\item \label{rq:cost}
\textbf{Does Skill Compilation reduce runtime token use?}

\item \label{rq:general}
\textbf{Does Skill Compilation improve end-to-end task success?}

\end{enumerate}

\paragraph{Experimental Setup}
\label{sec:experimental-setup}

We compare \textbf{SIGIL}, which executes compiled skill
harnesses while retaining model judgment, with
\textbf{\texttt{SKILL.md}}, which provides the original skills
to a ReAct-style agent for runtime interpretation and execution.
We evaluate adherence and runtime token use (RQ1--RQ2) on
11 public compliance-critical
skills~\citep{riskresponse_grc_skills,naik2026grcskills,kassis2026scientificagentskills},
where following the prescribed procedure is part of correctness
(Appendix~\ref{app:compliance-skills}).
We use three tasks per skill, yielding 33 tasks, and execute
each task three times per configuration.
We evaluate end-to-end task success (RQ3) on all tasks in
SkillsBench~\citep{skillsbench}.
Across all three RQs, we use Claude Opus 5, GLM-5.2, GPT-4o,
and Llama-3.1-8B as runtime models.
The compile-time models are Claude Opus 5, GLM-5.3,
and GPT-5.6-Sol.

\subsection{RQ1: Does Skill Compilation Improve Adherence to Skill Requirements?}
\label{sec:eval-rq1}

\begin{figure*}[t]
    \centering
    \begin{subfigure}[t]{0.40\textwidth}
        \centering
        \includegraphics[width=\linewidth]{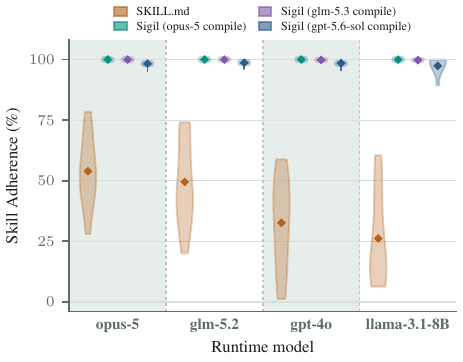}
        \caption{}
        \label{fig:sa-runtime}
    \end{subfigure}
    \hfill
    \begin{subfigure}[t]{0.58\textwidth}
        \centering
        \includegraphics[width=\linewidth]{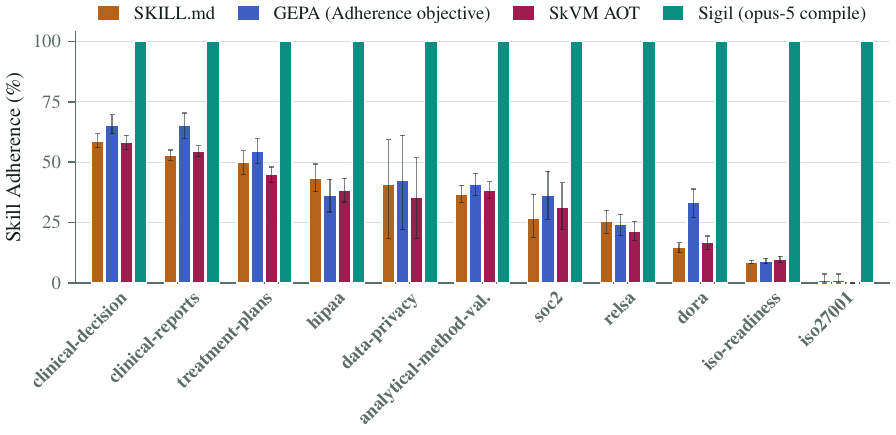}
        \caption{}
        \label{fig:sa-baselines}
    \end{subfigure}
    \caption{Skill Adherence.
    (a) Across four runtime models and three compile-time models,
    with each observation representing one task run.
    (b) Comparison with skill optimization and adaptation methods
    GEPA and SkVM, using GPT-4o at runtime and Claude Opus 5
    for SIGIL compilation.}
    \label{fig:rq1-adherence}
    \vspace{-0.5cm}
\end{figure*}

We used GPT-5 to extract each skill's requirements and ordering
constraints into a draft directed acyclic graph (DAG),
independently of the compiler's extraction.
Domain experts reviewed and revised the drafts against the
source skills (Appendix~\ref{app:expert-review}).
The reviewed DAGs served as fixed references throughout RQ1.

\subsubsection{Runtime Adherence}
\label{sec:runtime-adherence}

SIGIL compiled each skill once with each of Claude Opus 5,
GLM-5.3, and GPT-5.6-Sol, then reused each resulting harness
across all four runtime models.
\textbf{Skill Adherence (SA)} measures the fraction of applicable
reference requirements satisfied during execution, including
their ordering and flow constraints. GPT-5 judges every requirement
against the reference DAG using the task, available executable
structure, and execution trace.

Compilation substantially narrows the adherence gap across
runtime models (Figure~\ref{fig:sa-runtime}).
With direct execution of \texttt{SKILL.md}, mean SA ranges
from 26.16\% to 53.95\%. With Opus-compiled harnesses, all
four runtime models achieve 100\% measured mean SA.
The improvement also holds when Claude Opus 5 performs both
compilation and runtime execution, indicating that the benefit
is not limited to transferring procedures from a stronger
model to a smaller one.
Harnesses compiled with GLM-5.3 and GPT-5.6-Sol likewise
achieve 96.4--100\% measured mean SA across runtime models. We examine requirement preservation and structural
encoding as potential sources of the remaining variation in
\S~\ref{sub:reference-mandte-share}.

\vspace{-0.17cm}
\subsubsection{Comparison with Skill Optimization and Adaptation}

We include GEPA~\citep{agrawal2025gepa} and
SkVM~\citep{skillrt2026} to evaluate whether instruction
optimization and adaptation to the target runtime improve
skill adherence. We compare both with \texttt{SKILL.md}
and SIGIL using GPT-4o at runtime and the same SA metric.
GEPA was configured to optimize SA directly over 45 trials,
using GPT-5 reflection and feedback on individual requirements
and their ordering. SkVM used GPT-4.1 to adapt skills to the capabilities of
the target model and harness.

GEPA raises mean SA from 32.6\% to 37.2\%, but its gains vary
across skills (Figure~\ref{fig:sa-baselines}). It improves adherence
on DORA, while providing little benefit on ISO readiness and
ISO 27001, where all three baselines remain low.
SkVM achieves 31.7\% mean SA.
SIGIL reaches 100\% across all 11 skills, including those
where optimization and adaptation offer little benefit.
Under these configurations, optimizing or adapting skill
instructions provides limited adherence gains, supporting
explicit coordination of the recovered procedure.

\subsubsection{Compilation Fidelity and Mandate Realization}
\label{sub:reference-mandte-share}

We analyzed 33 AG-IR programs obtained by compiling all
11 skills with each of the three compile-time models.
We assessed these programs against each skill's expert-reviewed
reference mandates (mandatory requirements) combined across
its three evaluation tasks.
GPT-5 judge checked whether each reference mandate was encoded in
program structure, retained only as instructions for the runtime
model, or absent from AG-IR.
Structural encoding either enforces the requirement through
deterministic logic or requires a model call to perform the
prescribed step. For required model calls, the harness enforces
their occurrence under the applicable conditions, while their
outputs remain model-dependent.
We report the \textbf{share of reference mandates} in each
category, averaged across skills; structural and instruction-only
shares together give the total retained share.

\begin{figure*}[t]
    \centering
    \includegraphics[width=0.85\textwidth]{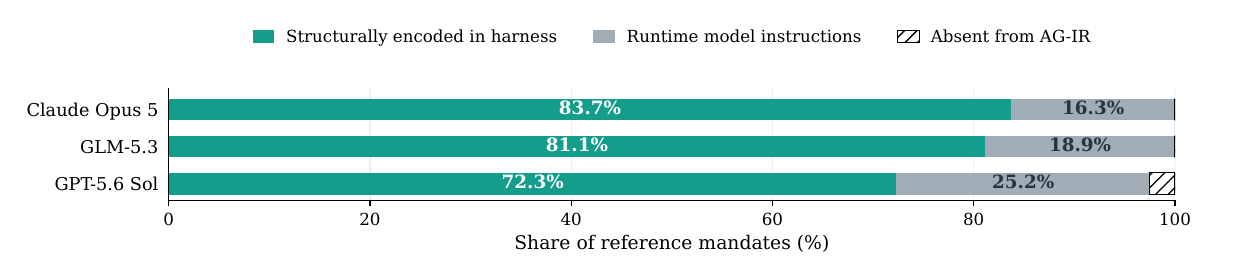}
    \caption{Reference mandate realization across compiler models.
    Structurally encoded requirements include deterministic enforcement
    and required model calls.}
    \label{fig:mandate-mechanisms}
\end{figure*}

Claude Opus 5 and GLM-5.3 preserve all reference mandates,
while GPT-5.6-Sol preserves 97.5\%.
Opus-compiled harnesses structurally encode the largest share
at 83.7\%, compared with 81.1\% for GLM-5.3 and 72.3\%
for GPT-5.6-Sol (Figure~\ref{fig:mandate-mechanisms}).
Compilers with lower structural encoding exhibit greater
adherence variation; GPT-5.6-Sol also omits reference mandates.
These findings suggest that the adherence gaps shown in
Figure~\ref{fig:sa-runtime} and discussed in
\S~\ref{sec:runtime-adherence} arise when compilation
omits requirements or the runtime model fails to follow
requirements retained only as instructions.

\begin{figure*}[t]
    \centering
    \begin{subfigure}[t]{0.47\textwidth}
        \centering
        \includegraphics[width=\linewidth]{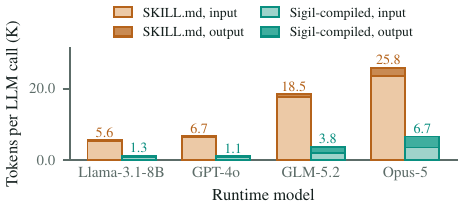}
        \caption{}
        \label{fig:rq3-tokens-per-call}
    \end{subfigure}
    \hfill
    \begin{subfigure}[t]{0.47\textwidth}
        \centering
        \includegraphics[width=\linewidth]{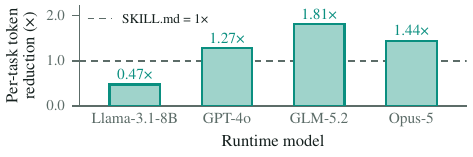}
        \caption{}
        \label{fig:rq3-token-reduction}
    \end{subfigure}
    \caption{Runtime token use, excluding compilation cost.
    (a) Mean input-plus-output tokens per model call.
    (b) Total \texttt{SKILL.md} tokens divided by SIGIL tokens.}
    \label{fig:runtime-efficiency}
    \vspace{-0.5cm}
\end{figure*}

\vspace{-0.3cm}
\subsection{RQ2: Does Skill Compilation Reduce Runtime Token Use?}
\label{sec:eval-rq2}

On the 11-skill suite used in \S~\ref{sec:eval-rq1}, we compared
mean total tokens per model call and total runtime tokens
between \texttt{SKILL.md} and SIGIL's Opus-compiled harnesses
across all four runtime models.
Figure~\ref{fig:rq3-tokens-per-call} shows that mean total tokens
per model call decrease by a factor of 4--6
across all four runtime models, primarily through lower
input-token use.
This suggests that scoping model calls helps reduce token use:
SIGIL's compiled harness supplies each call with inputs and
knowledge blocks relevant to its step, attaching relevant tools
when needed.
For three of these models, SIGIL uses 21--45\% fewer total
runtime tokens while achieving 100\% measured mean SA
(Figure~\ref{fig:rq3-token-reduction}).

Llama-3.1-8B instead uses 2.13 times as many total tokens
despite smaller calls, while mean SA rises from 26.2\% to 100\%.
Direct \texttt{SKILL.md} execution requires the model to track
required steps and decide when to stop. We observed it stop
before completing the procedure.
SIGIL's compiled harness completed those steps using additional
tokens. The baseline's lower token use therefore reflects
unfinished work.

\begin{figure}[t]
    \centering
    \begin{subfigure}[t]{0.29\textwidth}
        \centering
        \includegraphics[width=\linewidth]
            {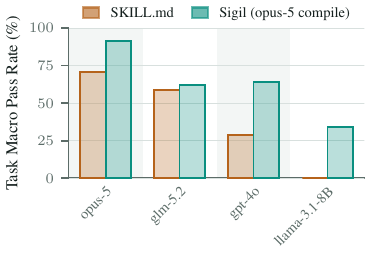}
        \caption{}
        \label{fig:task-macro-passed}
    \end{subfigure}
    \hfill
    \begin{subfigure}[t]{0.66\textwidth}
        \centering
        \includegraphics[width=\linewidth]
            {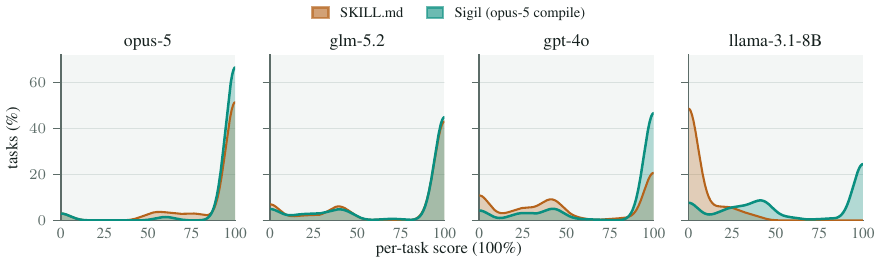}
        \caption{}
        \label{fig:skillsbench-distro}
    \end{subfigure}
    \caption{Task performance on SkillsBench.
    (a) Task Macro Pass Rate.
    (b) Per-task score distributions.}
    \label{fig:skillsbench}
    \vspace{-0.5cm}
\end{figure}

\subsection{RQ3: Does Skill Compilation Improve End-to-End Task Success?}
\label{sec:eval-rq3}

To test whether compilation improves task completion beyond
skill adherence, we evaluated all SkillsBench~\citep{skillsbench}
tasks. The benchmark supplies supporting skills and includes
tasks requiring reasoning beyond a single skill or coordination
across multiple skills.
Both approaches used an OpenHands agent to select skills,
supply inputs, and coordinate execution. The baseline provided
the agent with \texttt{SKILL.md}, while SIGIL provided compiled
harnesses through MCP.
We reused the same Opus-compiled harnesses across all four
runtime models. In each configuration, the same runtime model
powered the OpenHands agent and the model calls within the harnesses.
We report \textbf{Task Macro Pass Rate}, the equally weighted
mean across tasks of the fraction of SkillsBench's native
verifier checks passed, alongside per-task scores.

Figure~\ref{fig:task-macro-passed} shows that SIGIL improves
Task Macro Pass Rate across all four runtime models, with
increases of 20.8 percentage points for Claude Opus 5,
35.7 for GPT-4o, 34.1 for Llama-3.1-8B, and 3.5 for GLM-5.2.
GPT-4o and Llama-3.1-8B show the largest absolute gains.
Figure~\ref{fig:skillsbench-distro} shows that more tasks score
near 100\% with SIGIL for these two models and Claude Opus 5,
while GLM-5.2's score distributions remain broadly similar.
In execution traces, Opus delegated complete procedures to
compiled harnesses, while GPT-4o and GLM-5.2 sometimes
invoked them incrementally.
These observations suggest that compiled harnesses reduce
the procedural work handled by the OpenHands agent, while
task completion still depends on how it invokes and coordinates
the harnesses.

\vspace{-0.2cm}
\section{Related Work}
\label{sec:related}

GEPA optimizes instructions using execution feedback, while SkVM
adapts skills to model and harness capabilities
\citep{agrawal2025gepa,skillrt2026}.
SkCC emits framework-specific textual instructions that remain
subject to runtime interpretation~\citep{ouyang2026skcc}.
Voyager, Agent Workflow Memory, and related work acquire reusable
procedures from interaction and traces
\citep{wang2024voyager,wang2025awm,wang2026webxskill,
guo2026skilldisco,zhang2026workflowtoskill}.
These approaches improve skill acquisition, presentation, or
adaptation to the execution environment.
SIGIL takes an authored skill as its specification and translates
its procedural requirements into executable harness code.
Its objective is to preserve prescribed behavior, including
conditions, ordering, and prohibitions.

SkillSmith exposes executable operators, workflow graphs, and
guidance through boundary contracts interpreted by a shared
runtime, which handles selection, execution, and fallback
\citep{xu2026skillsmith}.
SIGIL instead makes the recovered procedure the target of code
generation: it represents source requirements and their execution
relationships in AIS and AG-IR, then deterministically lowers
the validated representation into Jac harness code.
Formal Skill defines executable skills through schemas, executors,
hooks, and state~\citep{zhang2026formalskill}; SIGIL addresses
deriving executable structure from an existing natural-language
skill. GuardAgent, ShieldAgent, and AgentSpec check or constrain
runtime behavior
\citep{xiang2025guardagent,chen2025shieldagent,wang2026agentspec}.
SIGIL encodes required operations and permitted transitions
directly in the generated procedure, while preserving semantic
model decisions and explicitly open subprocedures.

\vspace{-0.2cm}
\section{Conclusion}

We introduced Skill Compilation and SIGIL to encode authored
procedures in executable harnesses through a typed, validated
intermediate representation while preserving model judgment.
SIGIL achieves up to 100\% measured mean Skill Adherence across
11 compliance-critical skills and four runtime models, improves
SkillsBench task performance, and reduces total runtime tokens
for three models. These results support executable coordination
for more reliable and efficient skill execution across model
capabilities.


\subsection*{AI use statement}
We used generative AI tools to polish the manuscript, identify
related literature, review methodological descriptions, and
generate evaluation data and tasks. Language-model use in skill
compilation, reference extraction, and evaluation is described
in the corresponding sections. We reviewed the AI-assisted
materials and take responsibility for the final content,
data, and claims.




\bibliographystyle{iclr2027_conference}
\bibliography{references}



%

\appendix

\section{Compilation Details}
\label{app:method}

\subsection{Source Interpretation and AG-IR Construction}
\label{app:construction}

Algorithm~\ref{alg:skill-to-agir} summarizes compilation from a skill package
to validated AG-IR. The source package contains the skill text and its
associated resources. Interpretation produces procedural requirements $R$,
resolved executable bindings $B$, knowledge blocks $K$, and a source-role
map $H$. Requirements retain their source spans, modality, applicable
conditions, and prescribed relationships to other requirements. The
source-role map records how source content contributes to requirements,
bindings, and knowledge. A passage may serve several roles; source coverage
does not require converting every line into a procedural requirement.

Tool descriptions are resolved against supplied implementations or available
bindings. Descriptive text alone does not supply an executable operation.
Guidance and reference material are retained as knowledge and attached to the
operations that need them. A model-assisted coverage audit compares the
interpretation with the source package and identifies omitted requirements,
conditions, exceptions, ordering constraints, and prohibitions.

Given the recovered requirements and the Agent Instruction Set (AIS), the
compile-time model constructs instructions, control flow, typed carries, and
resource references. A requirement may be realized by several instructions
or by a constraint on their control or data flow. Subprocedures whose internal
steps are deliberately left open use \texttt{DELEGATE}. Composition produces
$A=\langle I,E,C,K,B,P\rangle$, where $P$ links the recovered requirements to
their source spans and realizations. The source-role map $H$ supports
interpretation and coverage auditing; it is not an additional AG-IR component.

\begin{algorithm}[t]
\caption{Model-assisted compilation from a skill package to AG-IR.}
\label{alg:skill-to-agir}
\small
\begin{algorithmic}[1]
\Require Skill $S$, resources $D$, available bindings $B_{\mathrm{avail}}$,
         AIS specification $\mathcal{O}$, compile-time model $M$,
         per-stage repair budget $J$
\Ensure Validated AG-IR $A$ or compilation failure
\State $X \gets (S,D)$
\State $(R,B,K,H) \gets \Call{InterpretSource}{M,X,B_{\mathrm{avail}}}$
\For{$j = 0,\ldots,J$}
    \State $G \gets \Call{CheckGroundingAndReferences}{R,B,K,H,X}$
    \State $Q \gets \Call{AuditSourceCoverage}{M,X,R,B,K,H}$
    \If{$G$ and $Q$ contain no unresolved issues}
        \State \textbf{break}
    \EndIf
    \If{$j = J$}
        \State \Return compilation failure
    \EndIf
    \State $(R,B,K,H) \gets
        \Call{RepairInterpretation}{M,X,R,B,K,H,G,Q}$
\EndFor
\State $Z \gets \emptyset$
\ForAll{$r \in R$}
    \State $Z[r] \gets \Call{RealizeRequirement}{M,r,R,\mathcal{O},B,K}$
\EndFor
\State $A \gets \Call{ComposeAGIR}{M,R,Z,\mathcal{O},B,K,H}$
\For{$j = 0,\ldots,J$}
    \State $E_{\mathrm{err}} \gets \Call{ValidateAGIR}{M,A,R,X,\mathcal{O}}$
    \If{$E_{\mathrm{err}} = \emptyset$}
        \State \Return $A$
    \EndIf
    \If{$j = J$ \textbf{or} $E_{\mathrm{err}}$ contains an unrecoverable failure}
        \State \Return compilation failure
    \EndIf
    \State $(E_{\mathrm{mech}},E_{\mathrm{sem}})
        \gets \Call{PartitionErrors}{E_{\mathrm{err}}}$
    \State $A \gets \Call{RepairMechanical}{A,E_{\mathrm{mech}}}$
    \If{$E_{\mathrm{sem}} \neq \emptyset$}
        \State $A \gets \Call{RepairAGIR}{M,A,R,X,E_{\mathrm{sem}}}$
    \EndIf
\EndFor
\end{algorithmic}
\end{algorithm}

\subsection{Validation and Repair}
\label{app:validation}

Before lowering, SIGIL checks the candidate against its recovered
requirements and the structural constraints of AG-IR
(Table~\ref{tab:gates}). These compilation checks use the source package and
the compiler's interpretation; they do not use the independent evaluation
references described in Appendix~\ref{app:expert-review}.

\begin{table}[t]
\caption{Pre-lowering validation checks. Reference validity, declared types,
and structural well-formedness are checked on the representation. Source
support, requirement realization, and semantic resource placement require
model judgment.}
\label{tab:gates}
\centering
\small
\begin{tabularx}{\linewidth}{@{}l >{\raggedright\arraybackslash}p{0.21\linewidth}
>{\raggedright\arraybackslash}X@{}}
\toprule
& \textbf{Check} & \textbf{Validation target} \\
\midrule
G1 & Source grounding
& Provenance references resolve to the source package, and represented
requirements are checked for support in their cited source context. \\
G2 & Requirement realization
& Recovered requirements are accounted for in the candidate; prescribed
conditions, ordering, and prohibitions are checked against the corresponding
operations and execution relationships. \\
G3 & Execution preservation
& Realizations preserve modality and the distinction among supplied
implementations, semantic computations, and deliberately open subprocedures,
including their associated bindings and knowledge. \\
G4 & Type and data flow
& Instruction inputs have compatible declared types and available values
on the execution paths where they are consumed. \\
G5 & Reference resolution and placement
& Control-flow, binding, knowledge, and provenance references resolve;
resources are associated with the operations that require them. \\
G6 & Binding and control-flow validity
& Executable operations have implementations or callable bindings, and
represented control-flow constructs are structurally well formed. \\
\bottomrule
\end{tabularx}
\end{table}

Mechanical defects are repaired directly when the correction is unambiguous
and does not require reinterpreting the prescribed procedure. Other defects
are returned to the compile-time model with the affected construct and
relevant source context. The repaired candidate is revalidated. In
Algorithm~\ref{alg:skill-to-agir}, each stage allows at most $J$ repair rounds
after its initial check; unresolved errors at the end of a stage cause
compilation failure.

Passing these checks means that no unresolved issue was identified by the
validator. It does not mechanically prove complete source interpretation or
correct semantic realization. In particular, the compiler's recovered
requirements and the independently reviewed evaluation requirements need
not coincide. Binding resolution and structural validity also do not
guarantee that an external operation succeeds or that a runtime model
returns a correct semantic result.

\subsection{Deterministic Lowering and Runtime Execution}
\label{app:lowering}

SIGIL lowers validated AG-IR to Jac using a fixed mapping from AIS
constructs to executable language constructs (Table~\ref{tab:lowering}).
Lowering does not call a model or reinterpret the source skill. The model
calls in the resulting program execute at runtime.

\begin{table}[t]
\caption{Lowering AIS and AG-IR constructs to Jac's
object-spatial programming (OSP) and model-invocation constructs.}
\label{tab:lowering}
\centering
\small
\renewcommand{\arraystretch}{1.12}
\begin{tabularx}{\linewidth}{@{}>{\raggedright\arraybackslash}p{0.32\linewidth}
>{\raggedright\arraybackslash}X@{}}
\toprule
\textbf{AG-IR construct} & \textbf{Jac/OSP realization} \\
\midrule

Instructions and control edges
& Operations executed by walker/node \texttt{can} abilities;
graph edges and \texttt{visit} statements encode traversal
between procedure steps. \\

\texttt{SELECT}, \texttt{EXTRACT},
\texttt{MUTATE}, \texttt{GENERATE}
& Typed functions implemented with \texttt{by llm()},
invoked from the corresponding abilities. \\

\texttt{DELEGATE}
& A tool-enabled \texttt{by llm(tools=[...])} invocation
executes the open subprocedure through a ReAct-style loop,
with the declared context, tools, and output contract. \\

\texttt{COMPUTE}
& Deterministic code in a walker/node \texttt{can} ability,
or a call to a supplied function from that ability. \\

\texttt{OBSERVE}, \texttt{ACT}
& Direct invocation of the bound tool or external operation
from the corresponding ability. \\

\texttt{BRANCH}
& Conditional \texttt{visit} statements or ordinary
\texttt{if}/\texttt{else} control flow over an available
predicate produced by an earlier operation. \\

\texttt{LOOP}
& Repeated traversal through a back edge using
\texttt{visit}, or a \texttt{for}/\texttt{while} loop
with the represented continuation condition. \\

\texttt{PARALLEL}
& \texttt{flow} launches independent procedure regions;
\texttt{wait} synchronizes their completion and results. \\

\texttt{CALL}
& A direct call to the named function, or
\texttt{spawn} of a walker implementing the named
subprocedure. \\

\texttt{RETURN}
& \texttt{report} emits the result and
\texttt{disengage} terminates walker execution;
function-based procedures use \texttt{return}. \\

Carries $C$
& Typed \texttt{has} fields on walkers or nodes,
accessed through \texttt{self} or \texttt{here},
and local variables within abilities. \\

Knowledge $K$
& Stored guidance in fields or variables, supplied
to the relevant model calls as arguments or
scoped context. \\

Bindings $B$
& References to callable functions, imported
implementations, or tool wrappers invoked by abilities. \\

\bottomrule
\end{tabularx}
\end{table}

The harness passes state between operations, invokes bound implementations,
and supplies semantic operations with their declared inputs and scoped
knowledge. Returned values determine subsequent execution according to
the represented control flow. Algorithm~\ref{alg:skill-to-agir} ends at
validated AG-IR; deterministic lowering and backend compilation follow this
stage, as shown in the main paper's compilation pipeline.

In the motivating \texttt{hipaa-compliance} skill used in the evaluation
suite~\citep{naik2026grcskills}, PHI classification remains a typed semantic
operation when the source provides classification guidance without an
executable classifier. If a BAA registry binding is available, the harness
obtains BAA status through that binding and guards transmission using the
returned status. Both results can depend on runtime input. The distinction
is whether the source and available resources supply an implementation or
leave semantic interpretation to the model.

Skills aggregated before compilation may produce one artifact. Separately
compiled artifacts are selected and coordinated by an outer runtime agent.
This coordination is distinct from \texttt{DELEGATE}, which preserves an
explicitly open subprocedure within a compiled skill.

\section{Compliance-Critical Skill Suite}
\label{app:compliance-skills}

Table~\ref{tab:compliance-skills} lists the 11 public skills used for
adherence (RQ1) and runtime token use (RQ2). The suite covers regulatory
compliance, scientific validation, and clinical workflows, where required
checks, ordering constraints, and prohibitions are part of correct execution
even when the final output appears acceptable. Three skills come from
Risk \& Response~\citep{riskresponse_grc_skills}, two from
Naik~\citep{naik2026grcskills}, and six from
Kassis et al.~\citep{kassis2026scientificagentskills}.

\begin{table}[t]
\caption{Compliance-critical skill suite used for RQ1 and RQ2.}
\label{tab:compliance-skills}
\centering
\small
\begin{tabularx}{\linewidth}{@{}>{\raggedright\arraybackslash}p{0.43\linewidth}
>{\raggedright\arraybackslash}X@{}}
\toprule
\textbf{Skill} & \textbf{Description and source} \\
\midrule
\texttt{soc2-system-description}
& SOC~2 system-description preparation and
validation~\citep{riskresponse_grc_skills}. \\

\texttt{data-privacy-expert}
& GDPR and CCPA privacy-compliance
workflows~\citep{riskresponse_grc_skills}. \\

\texttt{iso27001-internal-auditor}
& ISO~27001 internal-audit
workflows~\citep{riskresponse_grc_skills}. \\

\texttt{hipaa-compliance}
& HIPAA compliance assessment and
documentation~\citep{naik2026grcskills}. \\

\texttt{dora}
& DORA ICT-risk and resilience compliance
workflows~\citep{naik2026grcskills}. \\

\texttt{analytical-method-validation}
& Analytical-method validation against scientific
requirements~\citep{kassis2026scientificagentskills}. \\

\texttt{clinical-decision-support}
& Evidence-based clinical decision
support~\citep{kassis2026scientificagentskills}. \\

\texttt{iso-standards-readiness}
& Readiness assessment against applicable ISO
standards~\citep{kassis2026scientificagentskills}. \\

\texttt{relsa-severity-assessment}
& RELSA severity assessment and endpoint
forecasting~\citep{kassis2026scientificagentskills}. \\

\texttt{clinical-reports}
& Structured clinical reports and
documentation~\citep{kassis2026scientificagentskills}. \\

\texttt{treatment-plans}
& Structured clinical treatment
plans~\citep{kassis2026scientificagentskills}. \\
\bottomrule
\end{tabularx}
\end{table}

We use three tasks per skill, yielding 33 tasks, and execute each task three
times per configuration. SIGIL compiles each skill once with each of Claude
Opus~5, GLM-5.3, and GPT-5.6-Sol. Each resulting harness is reused across
Claude Opus~5, GLM-5.2, GPT-4o, and Llama-3.1-8B at runtime. The
\texttt{SKILL.md} baseline supplies the original skill to a ReAct-style agent.

The suite targets procedures for which adherence is part of correctness; it
is not intended to represent all public skills. SkillsBench complements
this evaluation by assessing end-to-end task performance in RQ3
(Appendix~\ref{app:skillsbench}).

\section{Reference Requirements and Expert Review}
\label{app:expert-review}

\subsection{Reference Construction}

GPT-5 extracts draft requirements and ordering constraints from the source
skills independently of the compiler's extraction. Domain experts review
and revise these drafts against the source skills and supporting resources.
The resulting directed acyclic graphs (DAGs) serve as fixed evaluation
references across compilation and runtime configurations.

Each reference identifies the mandates applicable to the evaluation task,
their conditions and exceptions, and their required ordering relationships.
Optional guidance is distinguished from mandatory behavior. Requirements
may specify individual actions, prohibitions, or a flow relationship such
as performing one action before another. This distinction is retained when
assigning failures to mandates during adherence scoring
(Appendix~\ref{app:adherence}).

\subsection{Expert Review Procedure}

Reviewers assess whether the references accurately reflect what the source
skill prescribes. They do not introduce obligations based solely on external
domain knowledge. Review materials comprise the source skill and supporting
resources, extracted requirements with source references, proposed ordering
constraints, and evaluation task descriptions.

Reviewers used the following questionnaire for each extracted requirement:
\begin{enumerate}
    \item \textbf{Source support.} Which source passage supports this
    requirement? Does the extracted wording preserve its meaning?
    \item \textbf{Mandatory or optional.} Does the source require, recommend,
    permit, or prohibit this behavior? Has optional guidance been treated
    as mandatory?
    \item \textbf{Conditions and exceptions.} When does the requirement
    apply? Are any conditions, exceptions, or stopping rules missing or
    incorrect?
    \item \textbf{Ordering and dependencies.} Must this operation precede
    or depend on another operation? Does the DAG capture the prescribed
    relationships without adding an order the source leaves open? Which
    mandates are attached to each ordering constraint?
    \item \textbf{Task applicability.} Does this requirement apply to the
    evaluation task? Which task details determine its applicability?
\end{enumerate}

After reviewing individual requirements, reviewers assess the complete DAG:
\begin{enumerate}
    \item \textbf{Completeness.} Are any required actions, checks,
    prohibitions, conditions, exceptions, or ordering relationships missing?
    \item \textbf{Consistency.} Are any requirements duplicated,
    contradictory, or unsupported? Does the DAG introduce restrictions
    absent from the source skill?
    \item \textbf{Readiness for approval.} Have identified issues been
    resolved in the reviewed version? Are any source ambiguities or
    applicability questions still unresolved?
\end{enumerate}

\paragraph{Review record and approval.}
Table~\ref{tab:dag-review-record} summarizes the review form used to document
expert review and approval. For each issue, the record identifies the affected requirement or ordering constraint,
its supporting source passage, the proposed correction, and the rationale.
Ambiguous source instructions are recorded for further review rather than
silently converted into mandatory obligations. The approval entry identifies
the reviewed DAG version and the reviewer's disposition; unresolved issues
remain explicit in the record.

\begin{table}[t]
\caption{Review form used for expert review and approval of reference DAGs.}
\label{tab:dag-review-record}
\centering
\small
\begin{tabularx}{\linewidth}{@{}>{\raggedright\arraybackslash}p{0.29\linewidth}
>{\raggedright\arraybackslash}X@{}}
\toprule
\textbf{Field} & \textbf{Recorded information} \\
\midrule
Reference identity & Skill, evaluation task, and DAG version. \\
Reviewer & Reviewer identifier and review date. \\
Affected element & Requirement or ordering-constraint identifier. \\
Source evidence & Supporting passage and source location. \\
Finding & Issue identified through the questionnaire. \\
Correction and rationale & Proposed change and explanation of source fidelity. \\
Resolution & Applied correction or remaining unresolved question. \\
Approval disposition & Approved, revision required, or unresolved; identify
the DAG version to which the disposition applies. \\
\bottomrule
\end{tabularx}
\end{table}

\section{Skill Adherence Evaluation}
\label{app:adherence}

\subsection{Evidence and Judgment}

Skill Adherence (SA) measures the fraction of applicable reference mandates
satisfied by an execution, including their required ordering and flow
conditions. GPT-5 judges every mandate against the reference DAG using
the task, available executable structure, and execution trace.
The same judging procedure applies to all evaluated methods and to
mandates represented structurally or retained as runtime instructions.

For each mandate, GPT-5 examines the corresponding operations and
execution relationships in the compiled AG-IR or available harness code,
together with execution evidence. It determines whether the prescribed
behavior and its applicable conditions were satisfied. Structural
representation provides evidence for this assessment but does not
receive automatic credit or bypass model judgment. Likewise, including
a requirement in a prompt does not establish that it was followed.

GPT-5 assesses branch choices against the task and reference conditions.
Incorrect branch choices fail the affected mandates. Mandates left
unsatisfied by a truncated execution also fail, even when their
implementations are present in the artifact.

\subsection{Ordering and Flow Violations}

GPT-5 checks both the required behaviors and their prescribed
relationships against the reference DAG. Performing individual actions
is insufficient when their required ordering is violated. A violated
ordering constraint fails every mandate attached to that constraint
in the reference.

If a reference mandate itself specifies the flow, such as ``perform $a$
before $b$,'' a violation fails that flow mandate. If the reference instead
contains separate mandates for $a$ and $b$, with their required ordering
attached to both, violating that order fails both mandates. Failure
assignment follows the reference's mandate representation; an ordering
violation does not introduce an additional mandate into the denominator.

Let $V_t$ denote the applicable mandates for task $t$, and let $F_t$ denote
its applicable reference ordering constraints. For each constraint
$e\in F_t$, let $L_t(e)\subseteq V_t$ contain the mandates attached to that
constraint. Let $\mathcal{B}_{t,u}\subseteq F_t$ contain the constraints
that GPT-5 judges violated in run $u$. The mandates failed by ordering are
\[
    \mathcal{F}^{\mathrm{order}}_{t,u}
    = \bigcup_{e\in\mathcal{B}_{t,u}} L_t(e).
\]
Each mandate receives one final pass/fail outcome, even when several
violations affect it.

\subsection{Prohibitions and Conditional Requirements}

GPT-5 judges each prohibition against the reference DAG, considering
the relevant execution path, any structural enforcement, and the
execution trace. An executable guard or excluded path provides evidence
of enforcement; a textual instruction alone does not. In every case,
GPT-5 determines whether the execution satisfied the prohibition.
An observed prohibited action fails the corresponding mandate.

For conditional requirements, GPT-5 assesses applicability using the
task and the source-specified conditions recorded in the reference.
An incorrect branch choice does not make an otherwise applicable
obligation inapplicable. Missing required behavior, broken flow,
and incorrect semantic decisions fail the affected mandates.
A required model call is assessed together with its result:
enforcing the call's occurrence does not establish that its output
satisfies the requirement.

\subsection{Score and Aggregation}

Let $s_{t,u}(m)\in\{0,1\}$ be the final GPT-5 judgment for applicable
mandate $m$ in run $u$ of task $t$, under the rules above. A mandate
passes only when its prescribed behavior and attached execution
constraints are judged satisfied. In particular,
$s_{t,u}(m)=0$ for every
$m\in\mathcal{F}^{\mathrm{order}}_{t,u}$. The run-level score is
\[
    \mathrm{SA}_{t,u}
    = \frac{1}{|V_t|}\sum_{m\in V_t}s_{t,u}(m).
\]
This aggregation counts the mandate-level judgments; it does not
assign separate automatic credit for structural enforcement.

For the compliance-critical suite, each skill has three tasks and
each task has three runs. We average run scores within each task,
then task scores within each skill, and finally across the 11 skills
with equal weight per skill. For a fixed configuration,
\[
    \overline{\mathrm{SA}}
    = \frac{1}{11}\sum_{k=1}^{11}
      \frac{1}{3}\sum_{t\in\mathcal{T}_k}
      \frac{1}{3}\sum_{u=1}^{3}\mathrm{SA}_{t,u},
\]
where $\mathcal{T}_k$ contains the three tasks for skill $k$.

\subsection{Optimization and Adaptation Baselines}

The GEPA and SkVM comparisons use GPT-4o at runtime, the same
reference requirements, and the same GPT-5 judging procedure
and SA aggregation as the other configurations.
GEPA optimizes SA over 45 trials, using GPT-5 reflection and
feedback on individual requirements and their ordering.
SkVM uses GPT-4.1 to adapt skills to the target model and harness.
SIGIL uses Opus-compiled harnesses for this comparison.

\section{Mandate Realization in Compiled Artifacts}
\label{app:mandate-realization}

\subsection{Artifacts and Reference Mandates}

We assess the 33 AG-IR programs produced by compiling each of the 11 skills
with Claude Opus~5, GLM-5.3, and GPT-5.6-Sol. Each skill's reference combines
the mandates from its three task references. These independently extracted,
expert-reviewed mandates are separate from the compiler's recovered
requirements.

The assessment examines how each reference mandate is represented in the
artifact. It inspects instructions, control and data flow, bindings, and
knowledge or instructions supplied to runtime model operations. GPT-5
assesses the correspondence between the reference mandate and its artifact
realization. This analysis concerns mandate representation in AG-IR rather
than the outcome of a particular execution.

\subsection{Realization Categories}

Each reference mandate is assigned to one of three categories:
\begin{description}
    \item[Structurally encoded.]
    The mandate is realized through executable operations or execution
    relationships. This includes deterministic enforcement and required
    model calls that perform a prescribed step under its applicable
    conditions. For a required model call, structure enforces the call's
    occurrence and placement; the semantic result remains model-dependent.
    Realization can involve several instructions or a relationship among
    them. Merely locating a node does not establish the required flow.

    \item[Runtime model instructions.]
    The mandate is retained in instructions supplied to a runtime model
    operation, but its satisfaction is not separately required by the
    represented structure. This category includes obligations embedded in
    a shared model operation and prohibitions expressed only in model
    instructions.

    \item[Absent from AG-IR.]
    The mandate is neither structurally realized nor retained in runtime
    model instructions in the assessed artifact.
\end{description}

For ordering mandates, structural realization requires the specified
execution relationship, not merely the presence of both operations. For
prohibitions, it requires structural exclusion or an executable gate that
blocks the prohibited behavior under the relevant conditions. Repeating a
prohibition in a dedicated model call does not by itself provide such an
enforcement.

\subsection{Aggregation and Scope}

For skill $k$ and compile-time model $c$, let $n_{k,c}^{(z)}$ be the number
of reference mandates in category $z$, and let $N_k$ be the number of
reference mandates for that skill. We compute the category share within
each skill and then average across skills:
\[
    p_c^{(z)}
    = \frac{1}{11}\sum_{k=1}^{11}
      \frac{n_{k,c}^{(z)}}{N_k}.
\]
The three shares sum to one. The structurally encoded and runtime-instruction
shares together give the retained share reported in the main paper.

This assessment characterizes preservation and realization of the reference
mandates. It does not by itself establish the absence of unsupported
additions, complete preservation of every source resource, or correctness
of runtime semantic outputs. One artifact per skill and compile-time model
does not estimate variation across repeated compilations.

\section{Runtime Token Evaluation}
\label{app:tokens}

RQ2 uses the same compliance-critical tasks as RQ1 and compares
\texttt{SKILL.md} with SIGIL's Opus-compiled harnesses across all four runtime
models. We report input-plus-output tokens per model call and total runtime
tokens. Compilation cost is excluded.

The total-token ratio shown in the main paper is
\[
    \rho
    = \frac{T_{\texttt{SKILL.md}}}{T_{\mathrm{SIGIL}}},
\]
where the numerator and denominator refer to total runtime tokens over the
corresponding evaluated workload. A ratio greater than one indicates lower
token use with SIGIL. The corresponding fractional reduction is
$1-1/\rho$. A smaller mean call does not necessarily imply lower total token
use, because the approaches can make different numbers of calls. Token
results are therefore reported alongside adherence on the same suite.

\section{SkillsBench Task Evaluation}
\label{app:skillsbench}

RQ3 evaluates all tasks in the SkillsBench benchmark used in the main paper.
The evaluated workload uses 87 tasks. The benchmark provides supporting skills and includes tasks requiring reasoning beyond a
single skill or coordination across multiple skills. Both approaches use
an OpenHands agent to select skills, supply inputs, and coordinate execution.
The baseline supplies the original \texttt{SKILL.md} files; SIGIL exposes
compiled harnesses through MCP.

The same Opus-compiled harnesses are reused across the four runtime models.
Within each configuration, the same runtime model powers both the outer
OpenHands agent and the model operations inside the harnesses.

For task $t$, let $v_t$ be the fraction of its native verifier checks passed.
The reported Task Macro Pass Rate is
\[
    \mathrm{TaskMacroPassRate}
    = \frac{1}{|\mathcal{T}|}\sum_{t\in\mathcal{T}}v_t,
\]
where $\mathcal{T}$ is the evaluated SkillsBench task set. This metric weights
tasks equally and credits partial verifier success. We also report the
distribution of per-task scores.

\end{document}